\documentclass{PoS}
\usepackage{tikz,graphicx,booktabs,multirow,array,microtype,mathtools,float,quark,braket}
\usepackage{graphicx}
\def\fm {\mathop{\hbox{fm}}}
\def\mev {\mathop{\hbox{MeV}}}

\usepackage{amsmath} % wonderful math package
\usepackage{dsfont}  % double strike font
\usepackage{bm}      % bold math

\def\beq{\begin{equation}}
\def\eeq{\end{equation}}
\def\beqs#1\eeqs{\beq\begin{split} #1 \end{split}\eeq}

\def\comment#1{}

\title{Resonance Parameters for the rho-meson from Lattice QCD}

\ShortTitle{Resonance Parameters for the rho-meson from Lattice QCD}

\author{\speaker{Dehua Guo} and Andrei Alexandru\\
        The George Washington University, Washington DC, USA\\
        E-mail:\email{guodehua@gwu.edu},~\email{aalexan@gwu.edu}}

\abstract{We present a calculation of the phase-shift for $\pi$-$\pi$ scattering in isospin-1, spin-1 channel in the elastic region. The mass and width of the rho resonance is extracted by fitting these phase-shifts. To vary the scattering momentum we employ asymmetric boxes. We use $N_f=2$ nHYP-smeared clover fermions and generate two sets of ensembles with pion masses about $315\mev$ and $227\mev$. To determine the phase shifts we compute the energy spectrum both for states at rest and boosted. We employ a variational analysis with interpolating fields including several $q\bar{q}$ and $\pi\pi$ interpolating fields with different scattering momenta.}

\FullConference{The 33rd International Symposium on Lattice Field Theory\\
		14 -18 July 2015\\
		Kobe International Conference Center, Kobe, Japan*}

\begin{document}
\section{Introduction}

Phase shifts for elastic scattering can be computed from the energies of two-particle states in a periodic box with the help of L\"{u}scher's method ~\cite{Luscher:1985dn,Luscher:1986pf,Luscher:1990ck,Luscher:1990ux}. In this study, we will focus on the $\rho(770)$ resonance that appears in isospin-1, spin-1 channel for $\pi$-$\pi$ scattering. 

To trace the phase shift through the resonance region, we vary the scattering momentum in small steps using a set of elongated boxes~\cite{Li:2003jn,Feng:2004ua,Li:2007ey,Pelissier:2011ib,Pelissier:2012pi}. The basic idea is that the particles scattering have a momentum of the order $\mathcal{O}(2\pi/ L)$, where $L$ is the size of the box. By increasing the direction of the lattice in one direction we access the relevant kinematic region while the cost of generating these ensembles is proportional with $L$ rather than $L^3$ as is the case for symmetric boxes. The geometry of the boxes used in this study was designed to produce momenta in the right kinematic region. Additional points in this region are determined by examining the spectrum two-hadron states with non-zero total momentum~\cite{Rummukainen:1995vs}.
We build interpolator fields with total momentum $\mathbf{P}= (0,0,0)$ and ones with $\mathbf{P}=(0,0,1)$ along the elongated direction. 

For each ensemble we extract phase-shifts for multiple center of mass energies by computing a tower of excited states in the relevant channel. To facilitate this we employ a variational method~\cite{Luscher:1990ck}, using a set of interpolator fields that overlap well with the lowest states in this channel. In our case we choose several quark-antiquark ($q\bar{q}$) interpolators and several $\pi$-$\pi$ interpolators with the momentum of the moving pions chosen to have a good overlap with the scattering states expected to appear in this region. The Wick contractions needed to compute the correlation matrix lead to a large number of diagrams. The most expensive ones are four-point diagrams that require basically the calculation of the all-to-all quark propagator. Stochastic methods can be used to compute these diagrams, but for large number of diagrams a better approach is to use LapH smearing~\cite{Peardon:2009gh}. We used LapH approach in this study.

We generated two sets of $N_f=2$ ensembles using nHYP-smeared clover fermions~\cite{Hasenfratz:2007rf} for two different sea quark masses: one set with pion mass $m_{\pi}\approx 315\mev$ and another set with $m_{\pi}\approx227\mev$. Each set has three ensembles with different elongation factor $\eta$, with geometry of the box $L\times L\times \eta L$. The details are listed in Table~\ref{tab:ensembles}. 
\begin{table}[b]
\begin{minipage}{0.49\textwidth}
\scalebox{0.7}{
\begin{tabular}{@{}*{7}{>{$}c<{$}}@{}}
\toprule
m_{\pi} & L/a & T/a  & a & \eta      & N_{cfg} & N_{vecs}\\
\midrule
315\mev &   24 & 48  & 0.121(1)    & 1.0   & 300   & 100     \\ 
        &      &     &             & 1.25  & 300   & 100     \\
        &      &     &             & 2.0   & 300   & 100     \\
%\rule{0pt}{3ex}    
227\mev &   24 & 64  & 0.1215(9)   & 1.0  & 400   & 100     \\
        &      &     &             & 1.17 & 400   & 100     \\
        &      &     &             & 1.33 & 400   & 100     \\
\bottomrule
\end{tabular}
}
\centering
\end{minipage}
\begin{minipage}{0.49\textwidth}
\scalebox{0.7}{
\begin{tabular}{c| l |c| l}
\toprule
\multicolumn{2}{c}{$O_h$} &\multicolumn{2}{|c}{$D_{4h}$}\\ 
\midrule
\multicolumn{1}{c|}{irreducible Reps.}  & \multicolumn{1}{c|}{$l$} & \multicolumn{1}{c|}{irreducible Reps.} &  \multicolumn{1}{c}{$l$} \\
\hline
$A_1$ & 0,4,6, ...                &  $A_1$& 0,2,3, ...\\
$A_2 $& 3,6, ...                  & $A_2$&1,3,4, ...\\
$F_1 $& 1,3,4,5,6, ...          &$B_1$ &2,3,4, ...\\
$F_2 $& 2,3,4,5,6, ...          & $B_2$&2,3,4, ...\\
$E$    & 2,4,5,6, ...               & $E$&1,2,3,4, ...\\
\bottomrule
\end{tabular}
}
\centering
\end{minipage}
\caption{(Left) Ensembles details: elongation $\eta$, lattice spacing $a$, number of configurations $N_{cfg}$ and number of eigenvectors $N_{vec}$ for LapH smearing. (Right) Angular momentum mixing for the irreps in $O_h$ and $D_{4h}$. }
\label{tab:ensembles}
\end{table}

\section{Symmetries and phase shift formula}

The eigenstates of the Hamiltonian in a finite box form degenerate multiplets that mix with each other under the symmetry transformations of the box. The multiplets correspond to invariant subspaces under the action of the symmetry group of the box. The action of the group in these invariant spaces is given by its irreducible representations ({\em irreps}). To connect the energy of the multiplets with the phase-shifts we need to identify the irrep of the eigenstate. To guarantee this we build interpolator fields with the appropriate transformation symmetry. 

The rotational symmetries of the cubic and elongated box are reduced from $SO(3)$ to $O_{h}$ and $D_{4h}$. The irreps of $O_{h}$ and $D_{4h}$ overlap with different irreps of $SO(3)$ labeled by the angular momentum $l$. The overlaps are shown in Table~\ref{tab:ensembles}. For $\rho$ resonance, we will focus on those irreps that mix with angular momentum $l=1$. In the $D_{4h}$ case, the relevant irreps are $A_2$ and $E$ irreps. $A_2$ is a one dimensional irrep and the lowest energy scattering states in this channel correspond to two-hadrons moving back-to-back in the elongated direction. $E$ is a two dimensional irrep and the lowest state corresponds to back-to-back momentum in the two transversal directions. Therefore, changing the elongation factor $\eta$ affects more the lowest eigenstates of the system in $A_2$ irrep. Hence, $A_2$ is the irrep we will focus in our study. We note that the $A_2$ irrep also mixes with $l=3$ and higher angular momentum states. In the $\rho$-resonance energy range the effect of these mixings is small and it is safe to neglect them.

To extract phase-shifts for additional center-of-mass energies, we also consider states with total momentum $\mathbf{P}=(0,0,1)$ in the elongated direction. For this choice the symmetry group remains the same, $D_{4h}$, since the relativistic expansion affects only the elongated direction. 

The L\"{u}scher's formula in the $A_2$ irrep for the states at rest is the following:
\begin{equation}
\cot\delta_1(k) = \mathcal{W}_{00} + \frac{2}{\sqrt{5}} \mathcal{W}_{20},
\end{equation}
where 
\begin{equation}
\mathcal{W}_{lm}(1,q^2,\eta) = \frac{\mathcal{Z}_{lm}(1,q^2,\eta)}{\eta\pi^{\frac{3}{2}} q^{l+1}}\,,\qquad q=\frac{k L}{2\pi}\,.
\label{eq:phaseshift_formula}
\end{equation}
The scattering momentum $k = \sqrt{\left(E/2\right)^2 - m_\pi^2}$ is extracted from the energy of the state $E$ and the mass of the pion.
The zeta function is defined as 
\begin{equation}
\mathcal{Z}_{lm}(s;q^2,\eta) =\sum_{\mathbf{n}\in P_\eta}\mathcal{Y}_{lm}(\mathbf{n})(\mathbf{n}^2-q^2)^{-s}\,,\qquad P_\eta=\{(n_1,n_2,n_3/\eta) | \mathbf{n}\in \mathbf{Z}^3 \} \,. 
\end{equation}
This formula can be extended to the boost case when the resonance has non-zero total momentum $\mathbf{P}=(0,0,1)$ by modifying the $\mathcal{W}$ and zeta function with the boost factor $\gamma$:
\beqs
\mathcal{W}_{lm}(1,q^2,\eta) &= \frac{\mathcal{Z}^{\mathbf{P}}_{lm}(1,q^2,\eta)}{\gamma\eta\pi^{\frac{3}{2}} q^{l+1}}\,,\quad E_{cm}=\sqrt{E^2-\mathbf{P}^2}\,,\quad \gamma=\frac{E}{E_{cm}} \,, \\
\mathcal{Z}^{\mathbf{\hat{P}}}_{lm}(s;q^2,\eta)&=\sum_{\mathbf{n}\in P_{\gamma\eta}}\mathcal{Y}_{lm}(\mathbf{n})(\mathbf{n}^2-q^2)^{-s}\,,\quad P_{\gamma\eta}=\{(n_1, n_2, (n_3+1/2)/(\gamma\eta) | \mathbf{n}\in \mathbf{Z}^3 \} \,.
\eeqs

\section{Interpolating fields}
 
In order to extract the energy spectrum from the correlation functions $\left< \mathcal{O}_i(t_f)\mathcal{O}_i^{\dagger}(t_i)\right>$ in $A_2$ irrep, we construct four $q\bar{q}$ interpolating fields and two $\pi\pi$ interpolators with different back-to-back momenta. All of these operators should have the same quantum number as the $\rho$ resonance. They $q\bar{q}$ interpolating fields are:
\begin{equation}
\rho^J(t_f)=\bar{u}(t_f)\Gamma_{t_f}A_{t_f}(\mathbf{p})d(t_f)\,,\quad \rho^{J\dagger}(t_i)=\bar{d}(t_i)\Gamma_{t_i}^{\dagger}A_{t_i}^{\dagger}(\mathbf{p})u(t_i) \,.
\end{equation}
where the $\Gamma$ matrices are listed in Table~\ref{tab:qbarq}.
\begin{table}[t]
\scalebox{0.8}{
\begin{tabular}{@{}*{5}{>{$}c<{$}}@{}}
\toprule
N  & \Gamma_{t_f} & A_{t_f} & \Gamma_{t_i}^{\dagger} & A_{t_i}^{\dagger} \\
\midrule
1  & \gamma_i & e^{i\mathbf{p}} & -\gamma_i & e^{-i\mathbf{p}}\\
2  & \gamma_4\gamma_i & e^{i\mathbf{p}} & \gamma_4\gamma_i & e^{-i\mathbf{p}} \\
3  & \gamma_i  & \nabla_j e^{i\mathbf{p}}\nabla_j &  \gamma_i & \nabla_j^{\dagger} e^{-i\mathbf{p}}\nabla_j^{\dagger} \\
4  & \frac{1}{2} & \{e^{i\mathbf{p}},\nabla_i\} & -\frac{1}{2} &\{e^{-i\mathbf{p}},\nabla_i\}\\
\bottomrule
\end{tabular}
}
\centering
\caption{$q\bar{q}$ operator lists.}
\label{tab:qbarq}
\end{table}
For the total momentum $\mathbf{P}=(0,0,0)$ case, the $\pi\pi$ interpolating field with back-to-back momentum along the elongated z-directions is:
\begin{equation}
\label{pipi100}
\pi\pi_{001}(\mathbf{p_1},\mathbf{p_2},t)=\frac{1}{\sqrt{2}}[\pi^+(\mathbf{p_1})\pi^-(\mathbf{p_2}) - \pi^+(\mathbf{p_2})\pi^-(\mathbf{p_1})]\,,\quad \mathbf{p}_1=(0,0,p_z) \ \ \mathbf{p}_2=-\mathbf{p}_1.
\end{equation}
The $\pi\pi$ operator with back-to-back momentum $\mathbf{p}_1=(0,1,1)$ and $\mathbf{p}_2=(0,-1,-1)$ generates a reducible four-dimensional subspace under the action of the symmetry group. The projection onto the $A_2$ irrep is:
\begin{equation}
\label{pipi110}
\pi\pi_{011}=\frac{1}{2}(\pi\pi(011)+\pi\pi(101)+\pi\pi(-101)+\pi\pi(0-11)).
\end{equation} 
A similar procedure is used to construct the $\pi\pi$ operators in a moving frame $\mathbf{P}\neq \mathbf{0}$. 
%Based on the variational method, to extract the excited energy states, we build up a $6\times6 $ correlation matrix:
%\begin{equation}
%C = \left(
%\begin{array}{ccc}
%C_{\rho^J \leftarrow \rho^{J'}}  & C_{\rho^J \leftarrow\pi\pi_{100}}  & C_{\rho^J\leftarrow\pi\pi_{110}}\\
%C_{\pi\pi_{100} \leftarrow \rho^{J'}} & C_{\pi\pi_{100}\leftarrow \pi\pi_{100}} & C_{\pi\pi_{100}\leftarrow \pi\pi_{110}}\\
%C_{\pi\pi_{110} \leftarrow \rho^{J'}} & C_{\pi\pi_{110} \leftarrow \pi\pi_{100}} & C_{\pi\pi_{110} \leftarrow \pi\pi_{110}}\\
%\end{array}\right),
%\end{equation}
%where $C_{\rho^J\leftarrow \rho^{J'}}$ is a $4\times 4$ block. 
%We use quark diagrams to simplify the Wick contraction procedure. For example, the $C_{\rho^J\leftarrow \rho^{J'}}$ and $C_{\rho_i\leftarrow\pi\pi}$ can be written as two points and three points diagrams:

We build up a $6\times6$ correlation matrix using four $q\bar{q}$ interpolators, $\pi\pi_{001}$ and $\pi\pi_{011}$. For $C_{\rho^J\leftarrow \rho^{J'}}$ and $C_{\rho_i\leftarrow\pi\pi}$ the correlation functions generated by the Wick contractions are the following
\begin{small}
\begin{equation}
\label{rho-rho}
C_{\rho_i \leftarrow\rho_j}=
 -\left<
\begin{array}{c}
\Gamma_{t_f}^J,(\mathbf{p},t_f)\\
\fish{->}\\
\Gamma_{t_i}^{J'\dagger}, (\mathbf{-p},t_i)
\end{array}
\right>\,,\quad 
C_{\rho_i\leftarrow\pi\pi}
= \left<
\begin{array}{c}
\uptriangle{-<}
\end{array}
 -\begin{array}{c}
  \uptriangle{->}
\end{array}\right>.
\end{equation}
\end{small}
Similarly, the $C_{\pi\pi\leftarrow \pi\pi}$ correlators are represented by the possible four points quark diagrams:
\begin{equation}
\label{pipi-pipi}
C_{\pi\pi\leftarrow\pi\pi}=
-\left<
\begin{array}{c}
\ssquare{->}
\end{array}
+
\begin{array}{c}
\ssquare{-<}
\end{array}
-
\begin{array}{c}
\doubletriangle{->}
\end{array}
-
\begin{array}{c}
\doubletriangle{-<}
\end{array}
+
\begin{array}{c}
\xstar{-<}
\end{array}
-
\begin{array}{c}
\doublefish{->}
\end{array}
\right>.
\end{equation}
The all-to-all propagators is required to evaluate the three-points and four-points quark diagrams. The Lapalacian Heaviside (LapH) method~\cite{Morningstar:2011ka} offers a way to efficiently evaluated the all-to-all {\em smeared} quark propagator. The quark fields in the interpolator fields are replaced with smeared quarks, leading to new interpolating fields with the same symmetry properties. The correlation functions for these interpolators require only the all-to-all smeared quark propagator. The smearing operator is constructed using the eigenvectors of the Laplacian operator up a cutoff $S_{\Lambda}(t) = \sum_{\lambda(t)}^{\Lambda}\Ket{\lambda(t)} \Bra{\lambda(t)}$. The radius of the smearing is controlled by the number of eigenvectors used in this expansion. We use 100 eigenvectors for all ensembles and the smearing radius is about $0.5\fm$.

\section{Phase shift and resonance parameters}

\begin{figure}[t]
\begin{minipage}{0.49\textwidth}
\includegraphics[scale=0.55]{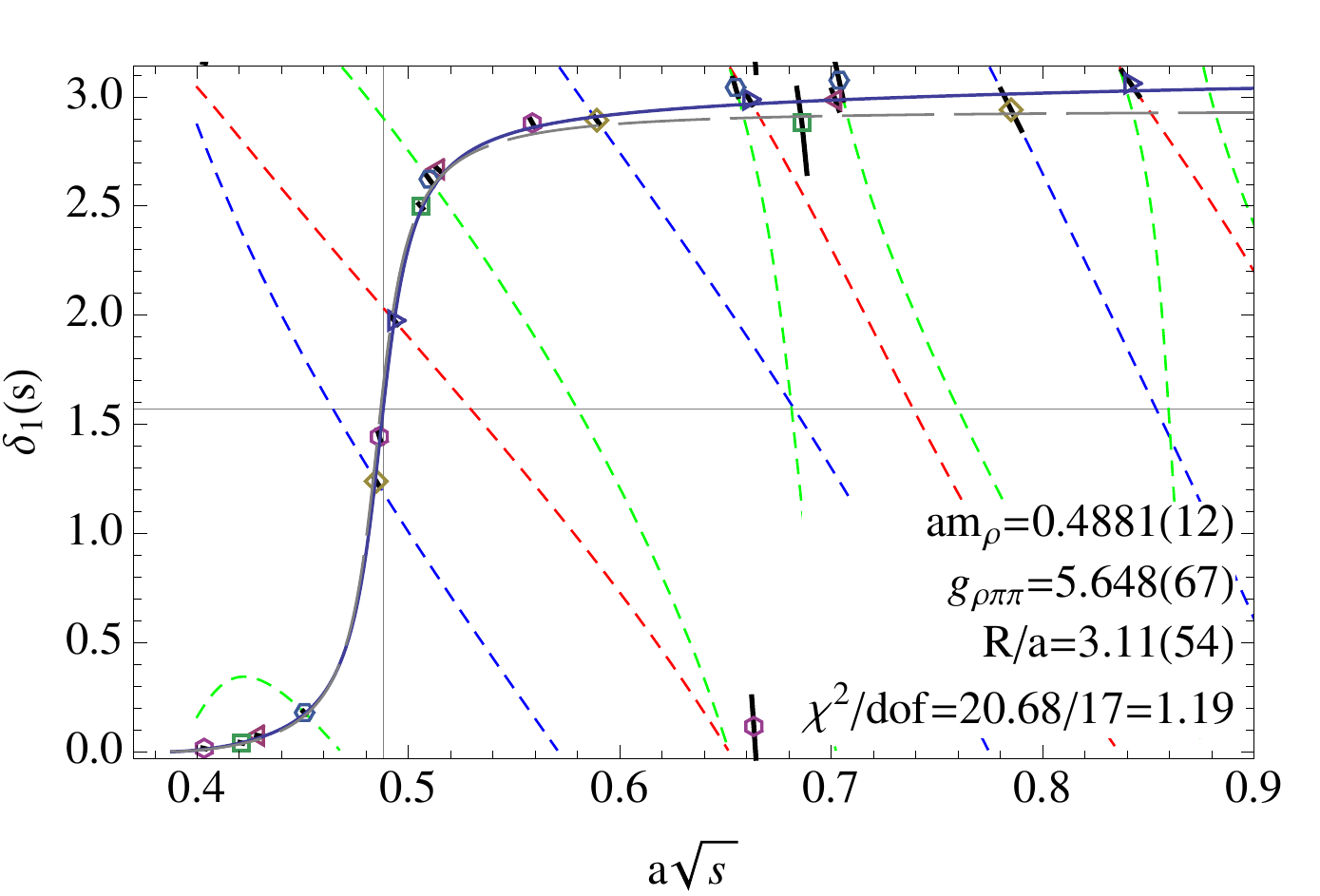}
\end{minipage}
\begin{minipage}{0.49\textwidth}
\includegraphics[scale=0.55]{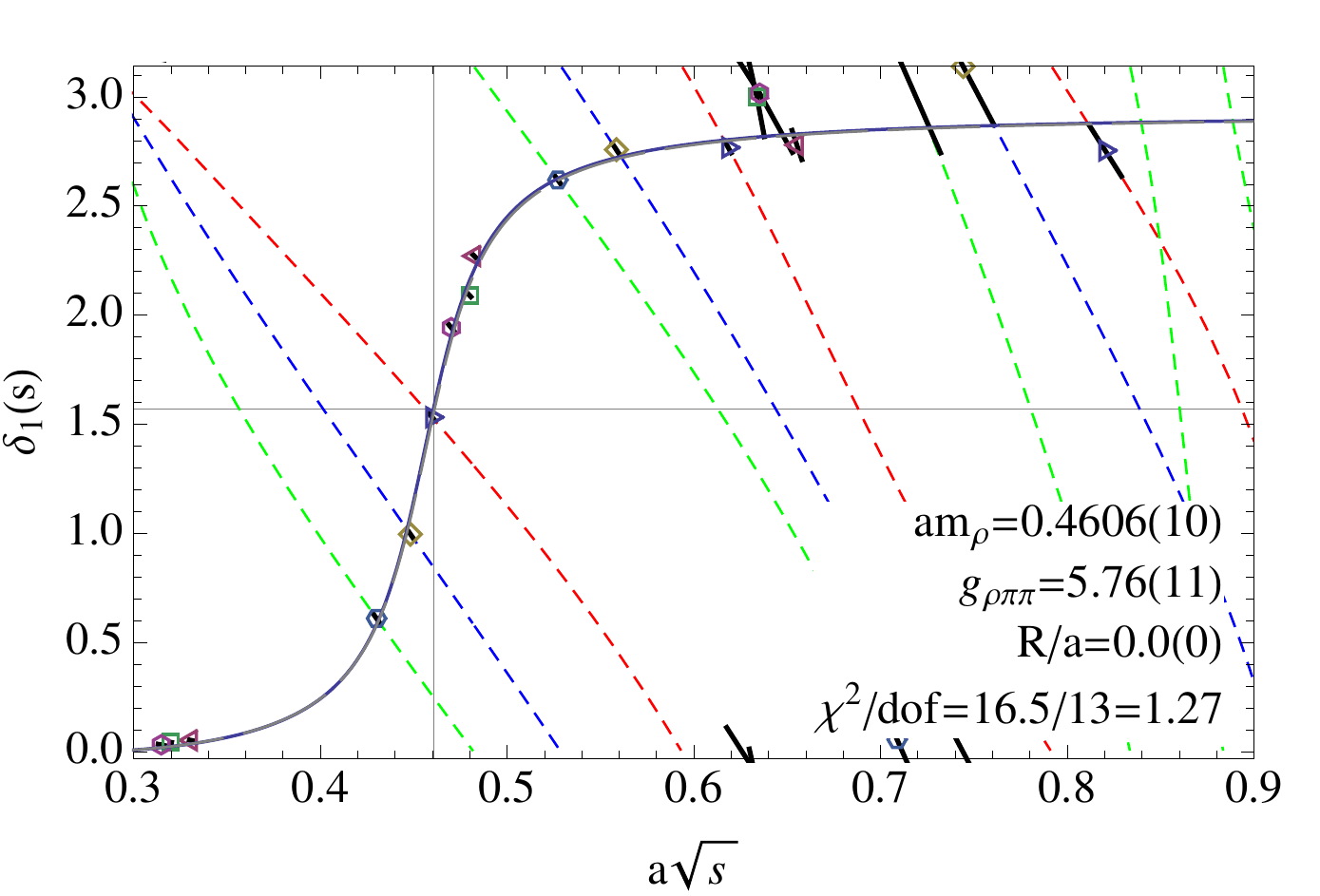}
\end{minipage}
\caption{(Left) $m_{\pi}\approx 315\mev$ phase-shift data fitted with both Breit Wigner form (dashed line) and modified form (solid line). (Right) $m_{\pi}\approx 227\mev$ ensembles data fitted with centrifugal barrier term. }
\label{fig:boost_phaseshift}
\end{figure}

We extract three energy levels for each ensemble both for $\mathbf{P}=(0,0,0)$ and  $\mathbf{P}=(0,0,1)$ states.
Using the formula in Eq.~\ref{eq:phaseshift_formula}, for each energy level, we compute the corresponding phase-shift. In Fig.~\ref{fig:boost_phaseshift} we show all the phase-shifts we computed.

To extract the resonance parameters we use the Breit-Wigner parametrization:
\beq
\label{eq:bw-param}
\tan \delta(E) = \frac{E\, \Gamma(E)}{m_\rho^2-E^2}
\quad \text{with} \quad \Gamma_\text{BW}(E) = \frac{g_{\rho\pi\pi}^2}{6\pi} 
\frac{p^3}{E^2}.
\eeq
We perform a correlated $\chi^2$ fit for the phase shift data and we find that the value of $\chi^2$ per degree of freedom is large indicating a poor quality fit. In Fig.~\ref{fig:boost_phaseshift} (left), the dashed line shows the fitting curve with Breit Wigner parameterization. The fit is bad because the Breit Wigner form doesn't describe well our data in high energy region. The same phenomenon was observed in experimental data~\cite{Estabrooks:1974vu} and in other lattice study~\cite{Dudek:2012xn}. One solution is to modify the Breit Wigner form decay width with a centrifugal barrier term~\cite{vonHippel} which makes the phase-shift value increase faster in the high energy region,
\beq
\Gamma(E_{\text{cm}})=\frac{g^2}{6\pi}\frac{p_{\text{cm}}^3}{E_{\text{cm}^2}}\frac{1+(p_RR)^2}{1+(p_{\text{cm}}R)^2}\,,
\label{eq:BW_centrifugal}
\eeq
where $R$ is a parameter called interaction radius. With the centrifugal barrier term, the high energy phase shift data can be well described by the modify Breit Wigner form in Fig.~\ref{fig:boost_phaseshift} (left). The interaction radius parameter has a reasonable value $R=0.376(65)\fm$. The phase-shifts for the $m_{\pi}\approx 227\mev$ ensembles are shown in Fig.~\ref{fig:boost_phaseshift}~(right). The quality of the fit is bad when we try to fit all points, even when including the centrifugal terms. In this case the points near the threshold cannot be captured by the fitting form. If we exclude the lowest three points from the fit, the rest of the data is fit well. Note that the fit also goes very near the lowest points, although it still misses them by a few sigmas.

\section{Comparison with other studies}

\begin{figure*}[t]
\begin{minipage}{0.50\textwidth}
\includegraphics[scale=0.8]{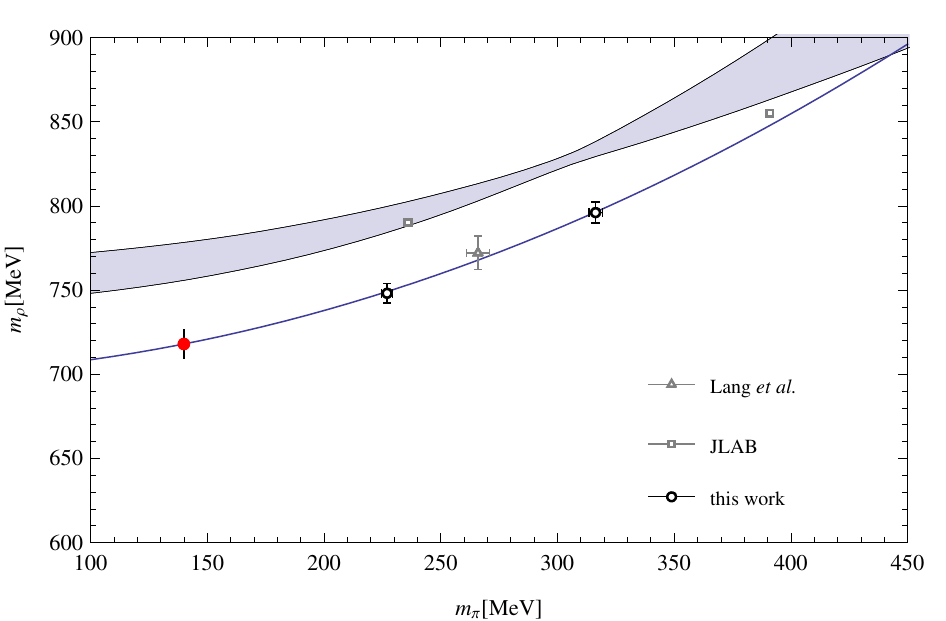}
\end{minipage}
\begin{minipage}{0.49\textwidth}
\includegraphics[scale=0.8]{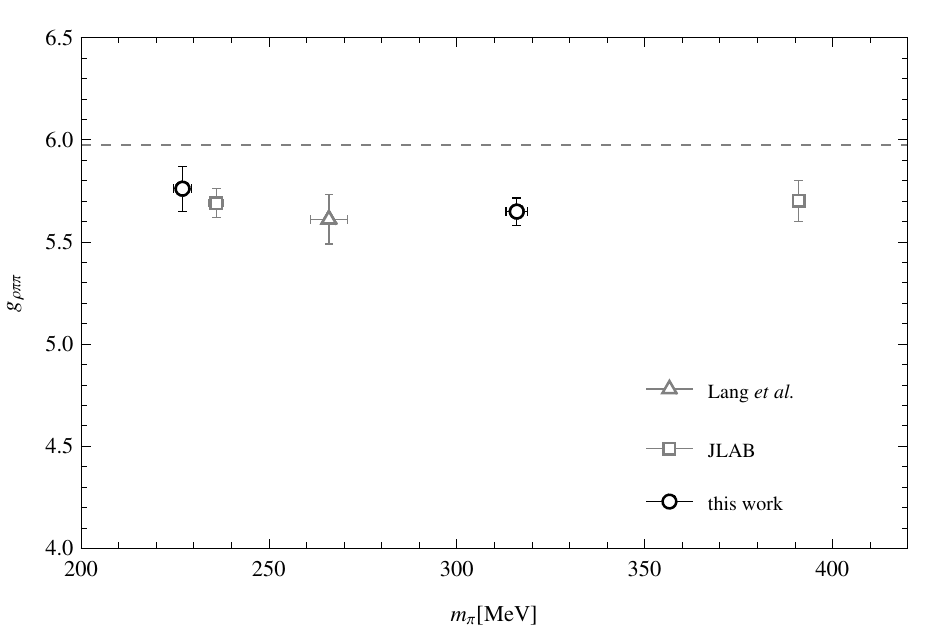}
\end{minipage}
\centering
\caption{Summary of recent lattice precision studies for $\rho$ resonance parameters~\cite{Lang:2011mn,Dudek:2012xn,Wilson:2015dqa}. The errorbars for our results include both statistical and the larger systematic error associated with determining the lattice spacing. The ones from JLAB include only statistical error. The blue curve is the $m_{\rho}$ extrapolation with $m_{\pi}^2$. The gray band presents the $m_{\rho}$ result study from unitarized chiral perturbation theory~\cite{Pelaez:2010fj} which includes the contribution from the strange quarks. Right panel shows $g_{\rho\pi\pi}$ versus $m_{\pi}$ from the same studies. The dashed line shows the value of $g_{\rho\pi\pi}$ as extracted from PDG~\cite{Agashe:2014kda} data.}
\label{fig:result_comparison}
\end{figure*}

We extracted the resonance parameters from the fits described in previous section. We extrapolate the resonance mass to the physical point using the expected relation between $m_\pi$ and $m_\rho$~\cite{Bruns:2004tj,Chen:2015tpa}: $m_{\rho}(m_{\pi}) = m_{\rho}^0 +c_1 m_{\pi}^2 + \mathcal{O}(m_{\pi}^3)$. We plot these results in Fig.~\ref{fig:result_comparison} together with recent precision studies for $\rho$ meson resonance from the lattice community. Note that our study and C. B. Lang et.al~\cite{Lang:2011mn} are performed in two mass-degenerated $N_f=2$ sea quarks without $K\bar{K}$ channel contribution. The results from JLAB group~\cite{Dudek:2012xn,Wilson:2015dqa} use $N_f=2+1$ sea quarks and the study at $m_{\pi}\approx 230\mev$ also includes the $K\bar{K}$ channel contribution.
Our extrapolation function $m_{\rho}(m_{\pi})$ goes through C. B. Lang et.al's result, which was extracted from an ensemble with significantly lower volume, which indicates that the finite volume effects are negligible. The extrapolation value at physical pion mass is $m_{\rho}=716(6)\mev$.

\section{Conclusion and outlook}

We performed a high precision study for rho-meson resonance parameters using nHYP smeared clover fermions at two dynamical pion masses: $m_{\pi}\approx 315\mev $ and $m_{\pi} \approx 227\mev$. We used elongated boxes and compute both rest-frame and boosted-frame energies to obtain the phase-shifts in the relevant kinematic region. Both quark-antiquark and multihadron operators are used in the variational analysis in order capture accurately the lowest three energy levels in each channel. The results for both pion masses show that the Breit Wigner form doesn't  describe well the phase-shift behavior away from the resonance region, but this can be addressed either by employing modifications of the fitting form or by restricting the fits to a narrow range around the resonance. The extrapolation for $m_{\rho}$ with $m_{\pi}$ based on our results is consistent with the other $N_f=2$ lattice study. The extrapolation value of $m_{\rho}$ at physical pion mass is about $60\mev$ lower than the physical $m_{\rho}$ value. This discrepancy is due to the missing $K\bar{K}$ channel contribution. A more detailed discussion will be included in an upcoming publication.

\section{Acknowledgments}

We would like to thank Craig Pelissier, Kevin Sykora, and Mike Lujan for generating some of the ensembles used in this study. The resources provided by the GWU IMPACT collaboration and GWU Colonial One cluster were used for this work. Both authors are supported by National Science Foundation CAREER grant PHY-1151648.

\bibliographystyle{JHEP-3}
\bibliography{ref}
%\begin{thebibliography}{99}
%bibitem{...}

%\end{thebibliography}

\end{document}